\documentstyle[prb,multicol,aps,epsf]{revtex}
\newcommand{\blong}{\ifpreprintsty
                   \else
                   \end{multicols}\vspace*{-3.5ex}{\tiny
                   \noindent\begin{tabular}[t]{c|}
                   \parbox{0.49\hsize}{~} \\ \hline \end{tabular}}
                   \fi}
\newcommand{\elong}{\ifpreprintsty
                   \else
                   {\tiny\hspace*{\fill}\begin{tabular}[t]{|c}\hline
                    \parbox{0.49\hsize}{~} \\
                    \end{tabular}}\vspace*{-2.5ex}\begin{multicols}{2}
                    \fi}
\def\be{\begin{equation}}
\def\ee{\end{equation}}
\def\bea{\begin{eqnarray}}
\def\eea{\end{eqnarray}}
\def\YBCO{YBa$_2$Cu$_3$O$_{7-\delta}$}
\def\Bi2212{Bi$_{2}$Sr$_{2}$CaCu$_{2}$O$_{8+\delta}$}

\def\ie{{\it i.e.} }
\def\etal{{\it et al.} }
\def\pg{{pseudogap} }
\def\bz{{Brillouin zone} } 
\def\cx{{$c$-axis} }
\begin{document}
\title{ $C$-axis Response of a High-$T_{c}$ Superconductor
with $D$-Density Wave Order}
\author{Wonkee Kim,$^{1}$ Jian-Xin Zhu,$^{2}$ J. P. Carbotte,$^{1}$
and C. S. Ting$^{2}$  }
\address{$^{1}$ Department of Physics and Astronomy, 
McMaster University, Hamilton,
Ontario, Canada, L8S~4M1\\
$^{2}$Texas Center for Superconductivity and Department of
Physics, University of Houston, Houston, TX 77204}
\maketitle
\begin{abstract}
The microscopic origin of the pseudogap state which exists in the underdoped
cuprates remains unknown. The $c$-axis properties in the pseudogap regime are
particularly anomalous. We use a recently proposed
model of a $d$-density wave which leads to staggered currents and the
doubling of the unit cell to investigate the $c$-axis kinetic energy and
the optical sum rule. The density of states expected in the model is
also considered.
\end{abstract}
\begin{multicols}{2}
\tighten

\section{introduction}

Pseudogap signatures in the underdoped regime of the 
high-$T_{c}$ superconductors
appear as intriguing features\cite{timusk} which could contain 
the key to a microscopic understanding of the physics of this important 
region of the phase diagram. It is in this region that proximity
to the Hubbard-Mott insulating phase would make one think that
the underlying normal state is most profoundly different from a usual metallic
Fermi liquid state. Many associated anomalous properties are
observed including important violation\cite{basov,katz}
of the $c$-axis optical sum rule
which related the missing area under the real part of the conductivity
in the superconducting state, 
as compared to the normal state,
to the corresponding $c$-axis superfluid
density. Above the superconducting critical temperature $T_{c}$, in the
\pg regime, important changes in the plasma frequency are also observed.
The preformed pair model\cite{emery} which envisions that electron pairs
are formed at the \pg transition temperature $(T^{*})$
with $T_{c}$ the temperature
at which phase coherence among the pairs
is established, has helped understand some of
these properties. In particular the \pg phase corresponds to
phase fluctuations with \pg energy the same as the pairing energy.
An alternative to the preformed pair model elaborated upon by
Chen \etal\cite{chen} envisions additional incoherent pair
excitations of finite momentum ${\bf q}$ which goes 
beyond the usual BCS formulation of superconductivity.
Very recently Chakravarty \etal\cite{chakravarty1}
have made a new proposal in which
the \pg is associated with the formation of a competing $d$-density
wave (DDW) state on a nested Fermi surface (FS). This DDW order
has important consequences such as staggered currents with associated
orbital magnetic moments which breaks parity and time-reversal
symmetry. Since the original proposal, several works have appeared
in which various properties associated with the DDW state have been
elaborated upon with the aim of testing
the validity of the model by comparing to experimental data.
\cite{han,tewari,chakravarty2}
ARPES data on the \pg have revealed it to have $d$-wave symmetry and
to exist at least in the underdoped regime. The DDW order
requires a nested FS and 
so should be most stable at half filling in the simplest of tight binding
bands, and by assumption the DDW order 
is taken to have $d$-wave symmetry with maximum gap
at $(\pi,0)$ and zero gap on the diagonals of the CuO$_{2}$ \bz.
It is therefore of considerable interest to understand how much more
of the known \pg physics such a model can explain.

Charge transport along the \cx has long been studied in the cuprates
and found to be anomalous. While for optimally doped \YBCO (YBCO)
the \cx DC resistivity tracks fairly well the linear in temperature
$(T)$ in-plane resistivity, above $T_{c}$, in the underdoped
case it shows a semiconductor-like increase\cite{cooper}
with decreasing $T$ and has a maximum at $T=T_{c}$.\cite{takenaka}
Also the estimated mean free path\cite{forro} becomes
shorter than the interlayer distance indication that Bloch transport
is unlikely to apply. A large decrease in the real part of the
conductivity $[\sigma_{1}(\omega)]$ is also observed at small
frequency $(\omega)$ starting at the \pg temperature $(T^{*})$.
The energy scale for the \pg is of the same order as that for the 
superconducting gap in YBa$_2$Cu$_3$O$_{6.6}$. 
At $T=0$,
the suppression of $\sigma_{1}(\omega)$ for frequencies $\omega$ below
the gap scale is very well developed. In fact the additional
spectral weight lost under
the $\sigma_{1}(\omega)$ curve on entering the superconducting state
is found to be smaller than the superfluid density determined at
$T\simeq0$ from the imaginary part of the conductivity.
The \cx sum rule is observed to be closer in value to a half
than to its conventional value\cite{fgt} of one. 
This indicates that there must be an 
important change in kinetic energy on going from the \pg state
at $T_{c}$ to the superconducting state at $T=0$. 
The observed value of the \cx sum rule of a half can be easily
understood within a preformed pair model,\cite{ioffe} as can the
important changes in the optical plasma frequencies which are
observed
in the temperature range
from $T_{c}$ to $T^{*}$ in the \pg state.
Within the preformed pair model these effects are the result of the phase 
fluctuations of the superconducting order parameter. 
In this paper we study these issues within the assumption that
the \pg state is due instead to the formation of the DDW order.

\section{formalism}

We begin with a 
phenomenological model Hamiltonian\cite{zhu} of the form:
$
H=-t\sum_{<i,j>\sigma}C^{+}_{i\sigma}C_{j\sigma}
-\mu\sum_{i\sigma}n_{i\sigma}
+J\sum_{<i,j>}
\left[{\bf S}_{i}\cdot{\bf S}_{j}-{1\over4}n_{i}n_{j}\right]\;,
$
where 
$C^{+}_{i\sigma}$ creates a spin $\sigma$ electron on the site $i$,
$t$ the in-plane hopping amplitude, $\mu$ the chemical potential, and
$n_{i}$ the occupation on the site $i$. ${\bf S}_{i}$ is the spin and
$J\;(>0)$ a strengh of an exchange interaction.
Based on the mean field approximation, defining 
$\Delta_{\delta}=-J<C_{i\downarrow}C_{i+\delta\uparrow}>$ and
making an ansatz
$(-1)^{i}W_{\delta}=-J<C^{+}_{i\sigma}C_{i+\delta,\sigma}>$,
we obtain the effective Hamiltonian for a combined superconducting
state with gap $\Delta_{\bf k}$ and \pg state (with \pg $W_{\bf k}$)
\blong
\begin{equation}
H_{MF}=\sum_{{\bf k}\sigma}(\epsilon_{\bf k}-\mu)C^{+}_{{\bf k}\sigma}
C_{{\bf k}\sigma}+\sum_{\bf k}
\left[\Delta_{\bf k}C^{+}_{{\bf k}\uparrow}C_{-{\bf k}\downarrow}
+h.c\right]
+\sum_{{\bf k}\sigma}iW_{\bf k}C^{+}_{{\bf k}\sigma}
C_{{\bf k}+{\bf Q}\sigma}
\end{equation}
\elong\noindent
where $\epsilon_{\bf k}=-2t[\cos(k_{x})+\cos(k_{y})]$,
$\Delta_{\bf k}=2\Delta_{0}[\cos(k_{x})-\cos(k_{y})]$, and
$W_{\bf k}=2W_{0}[\cos(k_{x})-\cos(k_{y})]$.
Here $\epsilon_{\bf k}$ is the electron energy dispersion for a simple
tight binding band with the FS at half filling coinciding with
the antiferromagnetic Brillouin zone. The superconducting gap amplitude 
is $\Delta_{0}$, and $W_{0}$ is the corresponding amplitude for the
\pg characterized by the commensurate wave vector ${\bf Q}=(\pi,\pi)$.
For our simple band, therefore,
$\epsilon_{{\bf k}+{\bf Q}}=-\epsilon_{\bf k}$,
$\Delta_{{\bf k}+{\bf Q}}=-\Delta_{\bf k}$, and
$W_{{\bf k}+{\bf Q}}=-W_{\bf k}$
and this symmetry can be used to limit
the summation on momentum to half the Brillouin zone. 

One can introduce an intersite
interaction $(V/2)\sum_{<i,j>,\sigma,\sigma'}n_{i\sigma}n_{j\sigma'}$ 
to allow for a difference in interaction to exist in the $d$-wave 
superconducting (DSC) and DDW channel:
$J\rightarrow J-V$ for DSC and
$J\rightarrow J+V$ for DDW. Note that the site-dependent factor
$(-1)^{i}$ in the ansatz made above ensures that nesting takes place for
half filling and that $W_{\delta}$ is pure-imaginary.

In the $4\times4$ matrix form,
$H_{MF}=\sum'_{{\bf k}}
{\hat C}^{+}_{{\bf k}}{\hat h}_{{\bf k}}{\hat C}_{{\bf k}}$,
where $\sum'_{{\bf k}}$ means a sum over half of Brillouin zone, and
${\hat C}^{+}_{{\bf k}}
=\Bigl(C^{+}_{{\bf k}\uparrow}, C_{-{\bf k}\downarrow},
C^{+}_{{\bf k}+{\bf Q}\uparrow}, C_{-{\bf k}-{\bf Q}\downarrow}\Bigr)$.
The $4\times4$ matrix ${\hat h}_{{\bf k}}$ is
\begin{equation}
{\hat h}_{{\bf k}}=
  \left(\begin{array}{cccc}
  \epsilon_{{\bf k}}-\mu&\Delta_{{\bf k}}&iW_{\bf k}&0\\
  \Delta_{{\bf k}}&-(\epsilon_{\bf k}-\mu)&0&iW_{\bf k}\\
  iW_{{\bf k}+{\bf Q}}&0&\epsilon_{{\bf k}+{\bf Q}}-\mu&
                              \Delta_{{\bf k}+{\bf Q}}\\
  0&iW_{{\bf k}+{\bf Q}}&\Delta_{{\bf k}+{\bf Q}}&
                              -(\epsilon_{{\bf k}+{\bf Q}}-\mu)
\end{array}\right)\;.
\end{equation}
This matrix defines our problem. We begin by studying \cx properties
and in particular the optical sum rule. 
To do this it is necessary to have
some model for the \cx charge transfer. For simplicity we will use
\be
H_{c}=\sum_{i\sigma}t_{\perp}\bigl[c^{+}_{i1\sigma}c_{i2\sigma}+
c^{+}_{i2\sigma}c_{i1\sigma}\bigr]\;,
\ee
where $t_{\perp}$ is the transfer integral for electron hopping between
two adjacent planes labeled by $1$ and $2$, respectively.
We could equally well have taken incoherent coupling.
The $c$-axis conductivity at frequency $\omega$ is
$\sigma_{c}(0,\omega)=(i/\omega)\Bigl[
\Pi_{ret}(0,\omega)-e^{2}d\langle H_{c}\rangle\Bigr]$,
where
$e$ and $d$ are an electron charge and an interlayer spacing, respectively,
and
\begin{equation}
\Pi(0,i\omega)=2e^{2}t^{2}_{\perp}dT
\sum_{\omega_{n}}\sum'_{\bf k}
{\mbox T}{\mbox r}
\Bigl[{\hat G}({\bf k},i\omega_{n})
{\hat G}({\bf k},i\omega_{n}+i\omega)
\Bigr],
\end{equation}
and
\begin{equation}
\langle H_{c}\rangle=2t^{2}_{\perp}T\sum_{\omega_{n}}
\sum'_{\bf k}
{\mbox T}{\mbox r}
\Bigl[{\hat M}{\hat G}({\bf k},i\omega_{n})
{\hat M}
{\hat G}({\bf k},i\omega_{n})\Bigr]\;.
\end{equation}
In the above equations, the $4\times4$ matrix Green function 
${\hat G}({\bf k},i\omega_{n})=
[i\omega_{n}-{\hat h}_{\bf k}]^{-1}$ and
${\hat M}=\left(\begin{array}{cc}
  {\hat\tau}_{3}&0\\
  0&{\hat\tau}_{3}
\end{array}\right)$, where ${\hat\tau}_{3}$ is a Pauli matrix.

The \cx superfluid density (or stiffness)
$(\rho_{s})$ in the superconducting state
is related to the limit as $\omega\rightarrow0$ of the
imaginary part of the conductivity. It is also related to the missing
spectral weight under the real part of the conductivity
on going from normal to superconducting state,
which we denote by $\Delta {\cal N}$ and define as
\be
\Delta {\cal N}=\int^{\omega_{c}}_{0^{+}}{}d\omega
\left[\sigma^{N}_{1}(\omega)-\sigma^{S}_{1}(\omega)\right]\;.
\ee
Here $\omega_{c}$ is a cut-off frequency of order the bandwidth, and  
N and S stand for normal and superconducting state, respectively.
Now, we have
\begin{eqnarray}
\rho_{s}=&&\Delta {\cal N}-4\pi e^{2}d
\Bigl[\langle H_{c}\rangle^{s}-\langle H_{c}\rangle^{n}\Bigr]
\nonumber\\
=&&4\pi\lim_{\omega\rightarrow 0}
[\omega\mbox{I}\mbox{m}\sigma_{c}(0,\omega)]\;.
\end{eqnarray}
From the above equations we obtain
a sum rule in terms of Green functions ${\hat G}$
and  ${\hat G}_{0}={\hat G}(\Delta=0)$ as follows:
\begin{equation}
{\Delta {\cal N}\over\rho_{s}}=
{1\over2}+
{1\over2}
{\sum'_{\omega,{\bf k}}\sum_{i,j}\alpha_{ij}
\left[G^{2}_{ij}-G^{2}_{0,ij}\right]
\over
\sum'_{\omega,{\bf k}}\sum_{i,j}\beta_{ij}
G^{2}_{ij}}\;,
\label{sum1}
\end{equation}
where 
\blong
\be
\sum_{i,j}\alpha_{ij}\left[G^{2}_{ij}-G^{2}_{0,ij}\right]=
\left(G^{2}_{11}-G^{2}_{0,11}\right)-\left(G^{2}_{13}-G^{2}_{0,13}\right)
-\left(G^{2}_{24}-G^{2}_{0,24}\right)+\left(G^{2}_{33}-G^{2}_{0,33}\right)
\ee
\elong\noindent
and
\be
\sum_{i,j}\beta_{ij}G^{2}_{ij}=G^{2}_{12}-G^{2}_{14}
-G^{2}_{23}+G^{2}_{34}\;.
\ee
Note that the superfluid density $\rho_{s}$ is given by
\begin{equation}
\rho_{s}={\cal C}T\sum_{\omega_{n}}\sum'_{\bf k}\sum_{i,j}
\beta_{ij}G^{2}_{ij}\;,
\end{equation}
where ${\cal C}=32\pi e^{2}t^{2}_{\perp}d$.
The equations above are easily generalized
for incoherent interlayer coupling.
As $T\rightarrow0$ (zero temperature limit),
\begin{equation}
\rho_{s}={\cal C}\sum'_{\bf k}\Delta^{2}_{\bf k}
\left[{1\over E^{3}_{1\bf k}}+{1\over E^{3}_{2\bf k}}\right]\;,
\label{super}
\end{equation}
where $E_{1\bf k}=\sqrt{(E_{0\bf k}-\mu)^{2}+\Delta^{2}_{\bf k}}$
and $E_{2\bf k}=\sqrt{(E_{0\bf k}+\mu)^{2}+\Delta^{2}_{\bf k}}$ with
$E_{0\bf k}=\sqrt{\epsilon^{2}_{\bf k}+W^{2}_{\bf k}}$. 

In the case when the DDW order $W_{\bf k}=0$, $E_{0\bf k}=|\epsilon_{\bf k}|$.
For $\epsilon_{\bf k}>0$, 
$E_{1\bf k}=\sqrt{(\epsilon_{\bf k}-\mu)^{2}+\Delta^{2}_{\bf k}}$
and $E_{2\bf k}=
\sqrt{(\epsilon_{{\bf k}+{\bf Q}}-\mu)^{2}+\Delta^{2}_{{\bf k}+{\bf Q}}}$
while for $\epsilon_{\bf k}<0$
$E_{1\bf k}=
\sqrt{(\epsilon_{{\bf k}+{\bf Q}}-\mu)^{2}+\Delta^{2}_{{\bf k}+{\bf Q}}}$
and $E_{2\bf k}=\sqrt{(\epsilon_{\bf k}-\mu)^{2}+\Delta^{2}_{\bf k}}$.
Here we have used the properties associated with the commemsurate wave vector
${\bf Q}=(\pi,\pi)$. Then, we obtain
\blong
\be
\rho_{s}={\cal C}\sum'_{\bf k}
\left\{
{\Delta^{2}_{\bf k}\over
\left[(\epsilon_{\bf k}-\mu)^{2}+\Delta^{2}_{\bf k}\right]^{3/2}}
+
{\Delta^{2}_{{\bf k}+{\bf Q}}\over
\left[(\epsilon_{{\bf k}+{\bf Q}}-\mu)^{2}
+\Delta^{2}_{{\bf k}+{\bf Q}}\right]^{3/2}}
\right\}
={\cal C}\sum_{\bf k}{\Delta^{2}_{\bf k}\over E^{3}_{\bf k}}\;,
\label{super0}
\ee
\elong\noindent
where the summation over ${\bf k}$ is unrestricted \ie is over
the entire Brillouin zone.
This is the usual expression of
$\rho_{s}$ with
$E_{\bf k}=\sqrt{(\epsilon_{\bf k}-\mu)^{2}+\Delta^{2}_{\bf k}}$
for the pure DSC case with no DDW order.

\section{$c$-axis response}

Based on experimental observations\cite{tallon} 
we choose the doping dependence of the DDW gap amplitude $W_{0}(x)$ to be
$W_{0}(x)=0.04(1-x/x_{c})$ in unit of $t$.
Here $x$ is doping and its critical value is
$x_{c}=0.2$, where the DDW gap disappears. 
Also we assume $\Delta_{0}(x)=0.02$ for $0.05<x<0.2$. 
A small variation of 
$\Delta_{0}(x)$ near the optimal doping $(x\simeq0.16)$, which is 
seen in the experiment,\cite{tallon} is not important
because our main consideration is focused on the underdoped regime of 
the cuprates. 
[See a later discussion and an inset of Fig.~2.] 
In Fig.~1,
we plot the calculated $c$-axis superfluid density vs 
the amplitude of the \pg $W_{0}(x)$
at $T=0$ assuming a constant chemical potential.
Such an assumption is of course not correct
since the chemical potential also changes with the dopping $x$; 
however, keeping it fixed will help us to
understand the physics. The solid curve is the $c$-axis 
superfluid density with the chemical potential
$\mu=-0.06$. The dashed curve is for a larger value of $\mu=-0.2$.
OD and UD at the top of Fig.~1 stand for overdoped and underdoped
regime, respectively.
One can see from Fig.~1 that when the absolute value of the chemical potential
$|\mu|$ is smaller than the \pg value of $W(x)=4W_{0}(x)$, which is its
maximum for a given doping $x$, then the superfluid density is reduced as
$W(x)$ in increased. This happens to be the case for most of the
doping values $x$ between $0.05$ and $0.2$ in the solid curve.
On the other hand, for larger values of $|\mu|$ the opposite holds
(dashed curve). To understand the physics behind this behavior
we return to the well known result embodied in Eq.~(\ref{super0}) 
which gives the superfluid density $\rho_{s0}$ for a pure $d$-wave
superconductor with no DDW gap \ie $W_{0}(x)=0$.
In this case assuming a cylindrical FS $\rho_{s0}={\cal C}N(0)/2$, where
$N(0)$ is the density of states at the FS. In this simple model,
$N(0)$ is constant throughout the band; however, if it did vary, then it would
be some appropriate energy average of $N(\omega)$ around the FS on the
scale of $\Delta_{0}$ that would replace it. We stress that for
$\rho_{s0}$ the size of the zero temperature superfluid density is
independent of the size of the superconducting gap $\Delta_{0}$ and
depends instead on normal state parameters. While  Eq.~(\ref{super})
has an explicit factor of $\Delta^{2}_{\bf k}$ in the numerator, this
does not translate directly into a reduced superfluid density at $T=0$
as $\Delta_{0}(x)$ decreases.

As the \pg develops it competes with the superconductivity but we can 
understand the trends seen in Fig.~1 for $\rho_{s}$ as a function of 
$W_{0}(x)$ by considering, first, the effect of $W_{0}(x)$ 
on the electronic density of states and then thinking of
switching on the superconductivity. In the inset of Fig.~1, we show the
quasiparticle density of states for the pure \pg state (\ie 
$\Delta_{0}(x)=0$). The solid curve is for $W_{0}=0.03$ \ie 
$x=0.05$ and the dotted curve is for $W_{0}=0.016$ \ie $x=0.12$.
We will return to a detail discussion of the density of states (DOS)
in a later section. For now suffice it to note that the logarithmic
singularities associated with the DOS are at
$|\mu|\pm 4W_{0}t/\sqrt{t^{2}+W^{2}_{0}}$. For a wide band with $t>>W_{0}$
they would be at $|\mu|\pm 4W_{0}$.
Now for $\mu=-0.2$ the Fermi surface FS$_{1}$ is localted away from the
DDW gap and the DOS aound the Fermi energy increases as doping ranges
from overdoped to underdoped \ie as $W_{0}(x)$ increases. Consequently,
$\rho_{s}$ at $T=0$ increase. However, when $\mu=-0.06$ the Fermi surface
FS$_{2}$ sits inside the DDW gap shown for most values of $x$ and 
an average of the DOS around FS$_{2}$ is reduced as $x$ becomes smaller
\ie $W_{0}(x)$ becomes larger. Now the value of the superfluid density
decreases as the underdoped regime is entered. Thus the opening of the 
\pg will decrease $\rho_{s}$ at $T=0$ only if $|\mu|$ is small enough
in the DDW model. For realistic values of $x$, however,
the chemical potential $\mu$ is not close to zero when it is
calculated self-consistently for the simple tight binding band structure
we have assumed and in the presence of a DDW order. 
While one can take empirical expression for the
doping dependence of superconducting gap and \pg, an arbitray choice of
$\mu(x)$ would not be consistent with our
model band structure and, therefore, is not allowed. We must calculate
$\mu$ from the filling.

From this discussion we conclude that the opening-up of a \pg
is not sufficient to lead to a large reduction in the \cx superfluid
density. On the other hand the interlayer hopping matrix
element $t_{\perp}(x)$ in the YBCO series is known to decrease almost
exponentially as the doping is decreased. This factor will
dominate over effects of the DDW order in consideration of the
suppression of the \cx superfluid density with doping.
These considerations imply that the doping dependence
of the \cx superfluid density is not a good quantity in which to study
the role of the DDW order. A better choice is the \cx optical
conductivity sum rule because it is independent of the
magnitude of the interlayer hopping amplitude and, therefore, of its
dependence on doping. 
We next turn to the
calculation of the \cx sum rule and focus on the issue of
whether or not the DDW model can describe
the experimental observations.\cite{basov}
In the YBCO system a conventional sum rule\cite{fgt}
of one is observed for the optimally doped case and 
of about a half for an underdoped sample.

For the sum rule calculation
we will make use of some experimental observation rather than
proceeding to a complete
self-consistent calculation. 
This is reasonable since the nature
of the interaction $J-V$ and $J+V$, which determine the size of the 
amplitude of DSC and DDW gap respectively in the formalism, is not known.
Assumptions on the variation of $\Delta_{0}$ and $W_{0}$ as functions of
$x$ based on empirical expressions correspond to specific
assumption about the unknown variation of the above interaction parameters.
However, in order to take into account
the band structure and doping we determine the chemical potential
as a function of doping and temperature by solving the filling equation
derived in the Bogoliubov-de Gennes formalism. 

In practice a more useful expression than Eq.~(\ref{sum1})
to calculate the sum rule, or the normalized missing spectral weight
(NMSW) $\Delta N/\rho_{s}$ is
\be
{\Delta N\over\rho_{s}}=1+{4\pi e^{2}d\over\rho_{s}}
\Bigl[
\langle H_{c}\rangle^{s}-\langle H_{c}\rangle^{n}\Bigr]\;.
\label{sum2}
\ee
Of course Eq.~(\ref{sum1}) is more direct if one wishes to 
understand how a sum rule of a half
can be obtained. This results when the Green's functions
in the second term of the right hand side cancel between superconducting
and \pg state as in the preform pair model.\cite{ioffe}
We see from Eq.~(\ref{sum2}) that
the kinetic energy difference
between the superconducting and pseudogap state divided by the superfluid
density determines the $c$-axis conductivity sum rule. 
Mathematical expressions for the kinetic energy and the superfluid
density are as follows:
\be
4\pi e^{2}d\langle H_{c}\rangle
=-{\cal C} T\sum_{\omega_{n}}\sum'_{\bf k}
{{\gamma_{1}+\gamma_{2}}\over\beta^{2}}\;,
\ee
and
\be
\rho_{s}={\cal C} T\sum_{\omega_{n}}\sum'_{\bf k}
{2\Delta^{2}_{\bf k}\alpha\over\beta^{2}}
\ee
where 
\bea
\gamma_{1}=&&
\left(\omega^{2}_{n}-E^{2}_{0\bf k}+\Delta^{2}_{\bf k}-\mu^{2}\right)
\left(\omega^{2}_{n}+E^{2}_{0\bf k}+\Delta^{2}_{\bf k}+\mu^{2}\right)^{2}
\\
\gamma_{2}=&&4\mu^{2}E^{2}_{0\bf k}
\left(E^{2}_{0\bf k}+3\omega^{2}_{n}
+3\Delta^{2}_{\bf k}+\mu^{2}\right)
\\
\beta=&&
\left(\omega^{2}_{n}+E^{2}_{0\bf k}+\Delta^{2}_{\bf k}+\mu^{2}\right)^{2}
-4\mu^{2}E^{2}_{0\bf k}
\\
\alpha=&&
\left(\omega^{2}_{n}+E^{2}_{0\bf k}+\Delta^{2}_{\bf k}+\mu^{2}\right)^{2}
+4\mu^{2}E^{2}_{0\bf k}
\eea
The kinetic energy
of each state should presumably be calculated 
at zero temperature because in some
notion of the theory of interlayer coupling, the kinetic energy difference
is associated with the condensation energy of 
superconductors.\cite{chakravarty} 
However,
Basov \etal have taken $T\simeq T_{c}$ since it is not easily to access
the pseudogap state at low $T$. In the calculation of the kinetic energy
we choose $T=0.01$ for the superconducting state and $T=0.1$ for the
pseudogap state. If $t=2000K$, then the working temperatures are
$20K$ and $200K$ in the superconducting and pseudogap state, respectively.

As we mentioned ealier, we determine self-consistently 
the chemical potential $\mu$ for a given
doping from the band structure as well as the given temperature
even though we adopt the experimental observation for $W_{0}$ and $\Delta_{0}$. 
The expression for
the filling $n$, in the self-consistent formalism, is written as:
\blong
\be
n=1+{1\over2}\sum_{\bf k}
\left[{{E_{0\bf k}+\mu}\over E_{2\bf k}}
\tanh\left({E_{2\bf k}\over2T}\right)
-{{E_{0\bf k}-\mu}\over E_{1\bf k}}
\tanh\left({E_{1\bf k}\over2T}\right)
\right]\;.
\label{filling}
\ee
\elong\noindent
In Fig.~2, we plot NMSW $(\Delta N/\rho_{s})$
as a function of doping $(x)$.
In the inset of Fig.~2 we reproduce some of the experimental phase
diagram obtained recently by Tallon and Loram\cite{tallon} which
shows critical temperature (plus sign $+$), \pg value $W_{0}$ (solid square)
and supercondicting gap (solid triangle) as functions of doping
for Y$_{0.8}$Ca$_{0.2}$Ba$_{2}$Cu$_{3}$O$_{7-\delta}$. Here the details are not
important and we take the data as typical of the cuprates.
We approximate, in unit of $t$,  
$W_{0}(x)=0.04(1-x/x_{c})$ and $\Delta_{0}(x)=0.02$ at $T=0.01$,
and $W_{0}(x)=0.025(1-x/x_{c})$ with $\Delta_{0}(x)=0$ at $T=0.1$.
As shown, in the underdoped regime NMSW is less than one while
it saturates to about one beyond the optimal doping. Different values
of $W_{0}(0)$ at $T=0.1$ do not change the overall behavior of NMSW vs $x$.
However, we did find that a different choice of working temperature
for the pseudogap state can change the doping dependence of NMSW
and, thus, the behavior shown in Fig.~2 is not robust.
It has been found that
a temperature dependence of NMSW for a $d$-wave superconductor
with a cylindrical FS was negligible.\cite{kim1}
The behaviors of NMSW in the tight binding band with a \pg are different.
While the DDW model gives the right trend for the doping dependence
of the \cx sum rule it does not fall to the value of a half observed in
the experimental work of Basov \etal\cite{basov}
We did not find a set of parameter with which this could be
achieved. In this regard we should mention that in the
preformed pair model it is argued that the result
$\Delta {\cal N}/\rho_{s}=1/2$ in the underdoped regime,\cite{ioffe}
follows simply because the Green's functions appearing in the
numerator of the second term in Eq.~(\ref{sum1}) (which are diagonal in
this case) are the same for superconducting and \pg state. In the preformed
pair model only the off-diagonal part of the Green's
functions differ in the two regime. It is this part which deals with
fluctuations of the phase coherence of the Cooper pairs.

Next we turn our attention to the \pg state above $T_{c}$ but below the \pg
temperature $T^{*}$. In this regime the optical spectral weight or 
plasma frequency is observed to vary with temperature.\cite{katz,ioffe,homes}
We reproduce in the inset of Fig.~4 the data on 
YBa$_2$Cu$_3$O$_{6.6}$ given in Ref.\cite{ioffe}
This implies that a change in \cx kinetic energy occurs as temperature is 
lowered from $T^{*}$, and in our model as the DDW gap grows. 

Within a self-consistent formalism, we calculate
the kinetic energy above $T_{c}$ and below $T^{*}$ in the regime when
DDW gap is included in the calculation, and we compare the
theoretical calculations with experimental results.\cite{ioffe} 
For simplicity,
we do not consider inhomogeneity in the system.
The self-consistent equations for $W_{\bf k}$ and the filling $n$ are:
\blong
\begin{equation}
W_{0}={V_{DDW}\over4}\sum_{\bf k}{W_{0}\eta^{2}_{\bf k}\over E_{0\bf k}}
\left[\tanh\left({{E_{0\bf k}+\mu}\over2T}\right)
+\tanh\left({{E_{0\bf k}-\mu}\over2T}\right)
\right]\;,
\end{equation}
\elong\noindent
and Eq.~(\ref{filling}) with $\Delta_{\bf k}=0$,
where $V_{DDW}$ is the DDW channel interaction and 
$\eta_{\bf k}=\cos(k_{x})-\cos(k_{y})$.
For given values of $V_{DDW}$ and $n$, $W_{0}(T)$ and $\mu(T)$ are 
self-consistently determined. 
In the calculation we choose the DDW channel interaction 
$V_{DDW}\simeq0.945$ and the values 
$n=1$, $0.99$, $0.97$, and $0.95$ 
for the filling. In fig.~3, we plot $W_{0}(T)$ 
versus $T$ in unit of $t$. In the inset we show
$\mu(T)$ as a function of $T$ 
for $n=0.95$. In this case, if we assume
$t=2000K$, then $T^{*}\simeq215K$. Since below $T_{c}$, which
is taken as $80K$, we would have to include an equation for $\Delta_{\bf k}$
as well in the self-consistent calculation and we have not,
we simply indicate $W_{0}(T)$ as a dashed line for $T<T_{c}$.
Using $W_{0}(T)$ and $\mu(T)$ for different $n$'s, we calculate
the normalized kinetic energy 
$\langle H_{c}\rangle(T)/\langle H_{c}\rangle(T^{*})$ for $T_{c}<T<T^{*}$.
An analytic expression for $\langle H_{c}\rangle$ is achievable
only if the magnitude of the chemical poteltial $|\mu|$ is
much less than 
a working temperature; namely, $|\mu|/T\ll1$. In this instance
it can be shown that
\blong
\begin{equation}
\langle H_{c}\rangle\simeq {t^{2}_{\perp}\over T}
\sum'_{\bf k}\left\{ \tanh^{2}\left({E_{0\bf k}\over T}\right)-1
+{1\over4}\left({\mu\over T}\right)^{2}\left[
\tanh^{2}\left({E_{0\bf k}\over T}\right)-1\right]
\left[3\tanh^{2}\left({E_{0\bf k}\over T}\right)-1\right]\right\}\;.
\end{equation}
\elong\noindent
The correction to the above equation is of order $(\mu/T)^{4}$.
In Fig.~4, we plot our numerical results for
the normalized kinetic energy for $T_{c}<T<T^{*}$. The curves are
labeled by the value of filling $n$ with $n=1.0$ half filling
and $n=0.95$ corresponding to a value of doping $x=0.05$. It is clear from 
the figure that near half filling the \cx kinetic energy and so the 
optical spectral weight does decrease very significantly as we go from
$T^{*}$ toward $T_{c}$. The reduction for half filling is by a factor
of $5$ more than necessary to agree with
experiment for YBa$_2$Cu$_3$O$_{6.6}$ shown in the inset of Fig.~4
which gives the experimental results for the square of plasma 
frequency as a function of $T$.\cite{ioffe}
However, as we move away from half filling, the initial 
reduction with decreasing temperature out of $T^{*}$ is rapidly
suppressed. Also at the lower temperatures considered, the optical 
spectral weight begins to increase again. The tight binding
band structure we have used in our calculations is at best genetic
for the oxides and may not apply in a quantitative sense for a particular
case. Nevertheless it is clear that the model of DDW order as an explanation 
of the physics of the \pg regime gives results for the change in
\cx kinetic energy with temperature which are rather sensitive to
the assumed parameters, for example to doping.

\section{density of states}

Renner \etal\cite{renner} have studied the evolution of the \pg features
in STM tunneling (SIN) in a series of \Bi2212 (Bi2212) as a function of doping
from underdoped to overdoped regime of the phase diagram.
Somewhat complimentary data are given in DeWilde \etal\cite{dewilde}
and in Miyakawa \etal\cite{miyakawa} where the authors concentrate
more on the hump and dip feature seen around twice the gap energy
in SIN and at three times in Josephson junctions.
Renner \etal\cite{renner} find a superconducting gap which is nearly
temperature independent up to $T_{c}$ at which point it merges smoothly
into a second gap like feature centered at the FS. They find the
\pg to be present both in underdoped and overdoped samples and that
its size scales in magnitude with the superconducting gap. This argues
for a common origin and these authors conclude that the data
is consistent with the idea of preformed pairs. Because they find that
the \pg is tide to the FS the data do not support a conventional
band structure explanation. On the other hand recent
intrinsic tunneling spectroscopic results\cite{krasnov1,krasnov2}
for mesa on Bi2212 single crystals (SIS) have revealed distinct features 
that can be associated with the superconducting gap and with 
the pseudogap. The temperature
dependence of the superconducting peak structure in the
dynamic conductance shows that it closes at $T_{c}$
while the hump structure which is associated with the
pseudogap is unaffected by the superconducting
transition. This was taken as evidence that the two phenomena are 
distinct and that the data do not
favor a preformed pair model. Later magnetic field studies showed that 
the structure identified with the 
pseudogap is insensitive to magnetic field while
the superconducting gap is strongly suppressed by the 
field,\cite{krasnov2,shibauchi}
which is taken as further confirmation of the identification made.
A feature of 
the data which is relevant to our work is that the superconducting gap 
corresponds to sharp peaks which are seen to grow even sharper 
with decreasing 
temperature and fall inside the \pg humps. 
The DDW model does not relate gap and pseudogap directly
and so is consistent with the above interpretation. 
In the self-consistent phenomenological model the two are related 
to $J-V$ and $J+V$ respectively and, therefore, they 
can be quite different in 
size and in variation with temperature.
On the other hand, 
the model does not give superconducting gap falling inside 
a larger pseudogap structure. This behavior is generic
to the model and has its origin in the fact that
the superconducting gap opens up at the FS while the
DDW gap is centered at the antiferromagnetic Brillouin zone boundary
as described below.

Although the density of states (DOS) ${N}(\omega)$ is
related to the in-plane dynamics, it is nevertheless
relevant to our present discussion because 
we have already seen that
the $c$-axis response
reflect in-plane dynamics. 
Employing standard manipulations,
it can be shown 
that the quasiparticle DOS ${N}(\omega)$ is:
\blong
\bea
{N}(\omega)=&&-2\sum'_{\bf k}\mbox{Im}\left[G^{R}_{11}+G^{R}_{33}\right]
\nonumber\\
=&&2\sum'_{\bf k}\left[u^{2}_{1\bf k}\delta(\omega-E_{1\bf k})
+v^{2}_{1\bf k}\delta(\omega+E_{1\bf k})
+u^{2}_{2\bf k}\delta(\omega-E_{2\bf k})
+v^{2}_{2\bf k}\delta(\omega+E_{2\bf k})\right]\;.
\label{dos}
\eea
\elong\noindent
Here  a factor of $2$ is for the summation over spin, 
$G^{R}$ is the retarded Green's function and the coherent factors are
$u^{2}_{1\bf k}={1\over2}\left[1+(E_{0\bf k}-\mu)/E_{1\bf k}\right]$
and $u^{2}_{2\bf k}={1\over2}\left[1-(E_{0\bf k}+\mu)/E_{2\bf k}\right]$
with $u^{2}_{i\bf k}+v^{2}_{i\bf k}=1$ $(i=1,2)$.
There are three limits of special interest:

i) When $W_{\bf k}=0$,
one can show that $N(\omega)$ reduces to DOS for the pure
DSC case as we already showed for the superfluid density.
In this instance,
for $\epsilon_{\bf k}>0$,
$E_{1\bf k}=E_{\bf k}$, $E_{2\bf k}=E_{{\bf k}+{\bf Q}}$,
$u^{2}_{1\bf k}={1\over2}\left[1+\xi_{\bf k}/E_{\bf k}\right]$,
and $u^{2}_{2\bf k}={1\over2}
\left[1+\xi_{{\bf k}+{\bf Q}}/E_{{\bf k}+{\bf Q}}\right]$,
where $\xi_{\bf k}=\epsilon_{\bf k}-\mu$ and
$E_{\bf k}=\sqrt{\xi^{2}_{\bf k}+\Delta^{2}_{\bf k}}$.
If $\epsilon_{\bf k}<0$, then $E_{1\bf k}=E_{{\bf k}+{\bf Q}}$,
$E_{2\bf k}=E_{\bf k}$, $u^{2}_{1\bf k}={1\over2}
\left[1+\xi_{{\bf k}+{\bf Q}}/E_{{\bf k}+{\bf Q}}\right]$, and
$u^{2}_{2\bf k}={1\over2}\left[1+\xi_{\bf k}/E_{\bf k}\right]$.
Therefore, we obtain
\blong
\bea
{N}(\omega)=&&\sum'_{\bf k}\Bigl\{\left[1+{\xi_{\bf k}/ E_{\bf k}}\right]
\delta(\omega-E_{\bf k})+
\left[1-{\xi_{\bf k}/ E_{\bf k}}\right]\delta(\omega+E_{\bf k})
\nonumber\\
+&&\left[1+{\xi_{{\bf k}+{\bf Q}}/ E_{{\bf k}+{\bf Q}}}\right]
\delta(\omega-E_{{\bf k}+{\bf Q}})
+\left[1-{\xi_{{\bf k}+{\bf Q}}/ E_{{\bf k}+{\bf Q}}}\right]
\delta(\omega+E_{{\bf k}+{\bf Q}})\Bigr\}
\nonumber\\
=&&\sum_{\bf k}\Bigl\{\left[1+{\xi_{\bf k}/ E_{\bf k}}\right]
\delta(\omega-E_{\bf k})+
\left[1-{\xi_{\bf k}/ E_{\bf k}}\right]\delta(\omega+E_{\bf k})\Bigr\}\;.
\eea
\elong

ii) When $\Delta_{\bf k}=0$, $E_{1\bf k}=|E_{0\bf k}-\mu|$ and
$E_{2\bf k}=|E_{0\bf k}+\mu|$. The coherent factors are now 
$u^{2}_{1\bf k}={1\over2}\left[1+\mbox{sgn}(E_{0\bf k}-\mu)\right]$
and 
$u^{2}_{2\bf k}={1\over2}\left[1-\mbox{sgn}(E_{0\bf k}+\mu)\right]$.
Simple algebra shows that the DOS for a pure DDW case is
\be
{N}(\omega)=\sum_{\bf k}
\left[\delta\left( \omega-(E_{0\bf k}-\mu) \right)
+\delta\left( \omega+(E_{0\bf k}+\mu) \right) \right]\;.
\ee

iii) When $\mu=0$, but, with nonzero DSC and DDW gap,
$E_{1\bf k}=E_{2\bf k}=\Xi_{\bf k}$, where $\Xi_{\bf k}=
\sqrt{\epsilon^{2}_{\bf k}+{\tilde\Delta}^{2}_{\bf k}}$ with
${\tilde\Delta}_{\bf k}=\sqrt{\Delta^{2}_{\bf k}+W^{2}_{\bf k}}$.
In this case the DOS becomes
\be
{N}(\omega)=\sum_{\bf k}\left[\delta(\omega-\Xi_{\bf k})+
\delta(\omega+\Xi_{\bf k})\right]\;.
\ee
Note that this DOS is
nothing but the DOS of a pure DSC case or
the DOS of a pure DDW case with a gap 
${\tilde\Delta}_{\bf k}$ for $\mu=0$.

For the solid curve in the top frame of Fig.~5, $W_{0}=0.04$ and 
$\Delta_{0}=0.02$ with the chemical potential $\mu=-0.25$ which is 
sufficiently large in absolute value that superconducting gap and \pg are
well separated, and there is distinct and characteristic structure
associated with each. The superconducting gap froms at the FS $(\omega=0)$
in the figure and the DDW gap at $\omega=|\mu|$. The position in energy
$\omega$ of the singularities corresponding to $4\Delta_{0}$ and
$4W_{0}$ are renormalized because both gap and \pg are present and
interfere, and because $\mu$ is not zero. But these renormalizations
are not large. For the \pg alone ($\Delta_{0}=0$ case)
there should be peaks of equal heights at $\omega\simeq
|\mu|\pm 4W_{0}$ \ie $\omega\simeq0.41$ and $0.09$. While
the uppermost peak is shifted upward only very slightly
by the presence of $\Delta_{0}$ to $0.418$, the second peak is shifted 
more importantly to $0.12$. If the \pg were zero, the superconducting 
peaks would be close but not quite at $\omega=\pm0.08$ because $\mu$ 
is not zero in this example. They are at $\omega=\pm0.076$.
Other than this small shift the gap and \pg structures are quite separate
in the solid curve. Note also from the general mathematical form 
Eq.~(\ref{dos}) for
the quasiparticle DOS that structures will fall symmetrically in energy
about $\omega=0$ but they are of different height. In particular we
note the small but clearly visible peak at $\omega=-0.12$.
The dot-dashed curve in the top frame of Fig.~5 is for comparison with the
solid curve and differs from it only through a different
value of $W_{0}=0.03$ instead of $0.04$. This change clearly
shifts the the two prominant \pg peaks leaving the superconducting gap
structure much less affected. It is clear from this comparison
that gap and \pg are pretty independent of each other although each is
somewhat affected by the presence of the other. If the chemical
potential is reduced to $\mu=-0.08$ as in the middle frame of Fig.~5, 
we can see that the competition between gap anf \pg becomes much more 
severe particularly in the region around the FS and the chemical 
potential which are now closer together. The position of the most prominant
upper peak falls at $\omega=0.173$ above the chemical potential ($|\mu|=0.08$).
This value is still close to $4W_{0}=0.16$. The third peak, which in the top 
frame was identified with the gap peak, falls at $\omega=0.036$ which is
a factor of $2$ lower than $4\Delta_{0}=0.08$ so that if we should identify
this with the gap,
we would have to conclude that it is strongly suppressed by
the presence of the DDW order. This is expected since both DSC and DDW order
compete for the available states in this region of energy. This
interpretaion finds further support in the fact that a change
in the value of gap $\Delta_{0}$ affects the position of this third peak
strongly while a change of the DDW gap $W_{0}$ does not. The position
of the second peak however changes significantly with a change
in $\Delta_{0}$ or $W_{0}$ so that this peak is a true mixture of both
gap and \pg.

In the bottom frame of Fig.~5, we show results for $\mu=0$ \ie at
half filling. In this case, as we have seen in the DOS,
gap and \pg  become a single entity ${\tilde\Delta}_{0}=
\sqrt{\Delta^{2}_{0}+W^{2}_{0}}=0.045$. A simple gap structure is obtained 
symmetric about $\omega=0$. 
In this case, the singularities are located at
$\pm 4{\tilde\Delta}_{0}t/\sqrt{t^{2}+{\tilde\Delta}^{2}_{0}}$.
It is clear from the trends exhibited
in Fig.~5 that the DDW model does not allow for a separate well identified
set of two gaps with superconducting gap falling inside a larger \pg
about $\omega=0$.
This could only happen if gap and
\pg where both pinned to the FS. In the DDW model only the superconducting
gap is pinned to the FS while the DDW gap opens up at the energy of
the antiferromagnetic Brillouin zone. Of course the intrinsic tunneling
experiments do not measure the DOS directly as they involve SIS junctions.
Our calculations can be compared more directly to the SIN data of 
Renner \etal\cite{renner} which do not presently show the evolution with
doping expected in a DDW model.

\section{conclusions}

As a possible candidate model for the \pg state seen in the
underdoped regime of the high-$T_{c}$ cuprates, we have studied the effects
of the formation of $D$-density wave (DDW) order on several properties.
While it is found that the value of zero temperature superfluid
density $(\rho_{s})$ is impacted by the opening-up
of a DDW gap at the antiferromagnetic Brillouin zone, its effect can increase 
as well as decrease the value of $\rho_{s}$. If the chemical potential
is small and near zero, the effective normal state density of electronic
states around the Fermi surface (FS) is on average reduced when the \pg
is increase by going to a more underdoped case. This leads to a reduction in
$\rho_{s}$. But if the chemical potential falls well above the \pg
energy, the opposite holds. In any case these effects are not dominant
and are masked by the near exponential reduction
in interlayer transfer matrix element known to exist in
{\YBCO} as one goes from overdoped (where $t_{\perp}\sim30-40$mev) to
underdoped case (where $t_{\perp}$ can be a fraction of a meV).\cite{cooper}

A quantity which does not depend explicitly on $t_{\perp}$ and its
variation with doping is the \cx optical sum rule, which is known
to be anomalous in underdoped cuprates while it takes on its
conventional value of one in optimally doped \YBCO. In this case
we find that opening a DDW gap will affect the value of the ratio of
the missing area under the difference between the real part of the
optical conductivity in \pg and superconducting state.
When it is normalized to the value
of the \cx superfluid density, which is determined from the
zero frequency limit of the imaginary part of the conductivity, it is found
to be close to one in the overdoped regime and decrease towards a half
in the underdoped case. In the case presented, the minimum value obtained
was a little less  than $0.8$ but the calculations show that
the size of the reduction depends on details of band structure, on
filling, and on size of the gap and \pg as well as 
on the reference temperature used. 
A reduction can occur but its size is not a robust feature of the model.

We also considered how the area under the real part of the
conductivity which gives the plasma frequency (or optical spectral
weight) is affected by the introduction of the DDW order. It is found that
near half filling, the optical spectral weight is substantially
reduced as the temperature ($T$)
is decreased from the \pg temperature ($T^{*}$)
towards
the superconducting transition temperature ($T_{c}$). 
Agreement with experiment is possible near half filling.
When the doping
is increased, however, away from half filling, the plasma frequency is
found to have a minimum in the region $T_{c}<T<T^{*}$. This behavior
is different from that expected
on the bases of the preformed pair model where the
phase fluctuations are the cause of the variation in optical spectral
weight which only decreases with decreasing $T$.

To acheive some physical understanding of our \cx result we found it
useful to introduce and consider the modifications
brought to the in-plane quasiparticle density of states
through the growth of DDW order. Because a generic feature of the model is that
superconducting gap opens at the Fermi surface while the DDW gap 
opens instead at the magnetic Brillouin zone, the resulting density of states 
$N(\omega)$ never shows superconducting gap features distinct from
\pg feature which are also positioned inside the \pg humps as reported 
in a recent experiment. The prediction is that when the 
chemical potential ($\mu$) is
large enough, gap and \pg features are separated by $|\mu|$ and are
quite distinct. 
When $\mu=0$, gap and \pg can no longer be distinguished
and the square root of the sum of the squares plays the role
of a single gap instead, reminiscent of the preformed pair model.

\begin{center}
{\bf AKNOWLEDGMENTS}
\end{center}

W.K aknowledges A. Ghosal for helpful discussions.
Research was supported in part by the National Sciences and Engineering
Research Council of Canada and by the Canadian Institute for
Advanced Research (W.K and J.P.C) and by the Texas Center  
for Superconductivity at the University of
Houston through the State of
Texas (J.Z and C.S.T).

\end{multicols}

\begin{figure}

\caption{
The \cx superfluid density $\rho_{s}$ divided by its maximum value
as a function of the value of the \pg amplitude $W_{0}(x)$ which depends on
the doping $x$. In the solid curve the chemical potential $\mu$
is fixed at $-0.06$ and in the dashed curve at $-0.2$ in a unit of $t$. 
The inset gives
the density of quasiparticle states (DOS) in a pure DDW state. The solid
curve is for $W_{0}=0.03$ and the dotted for $0.016$. Two possible
position for the Fermi surface (FS$_{1}$ and 
FS$_{2}$) are indicated by vertical lines. OD and UD on the top
stand for 
overdoped regime and underdoped regime, respectively.
}
\vskip 0.5cm

\caption{
The value of the \cx sum rule $\Delta{\cal N}/\rho_{s}$ as a function of
doping $x$. The changes reflect the opening  of the DDW gap as the
underdoped regimed is entered. The inset shows a typical phase diagram
for the cuprates from Ref.\cite{tallon} with the superconducting dome 
(plus sign $+$), \pg (solid square) and superconducting gap (solid triangle)
in unit of $[K]$.
Note that for both superconducting gap and \pg their maximum on the FS is
four times $\Delta_{0}$ and $W_{0}$, respectively.
}
\vskip 0.5cm

\caption{
The value of the DDW gap $W_{0}(T)$ as a function of temperature ($T$)
determined self-consistently for a given
value of parameters ($n=0.95$ and $V_{DDW}\simeq0.945$)
as described in the text. The inset
shows the corresponding chemical potential $\mu$ as a function of $T$.
}
\vskip 0.5cm

\caption{The \cx kinetic divided by its value at $T^{*}$ as a function of
temperature ($T$) in the range $T_{c}<T<T^{*}$. The curves are labeled
by the values of the filling $n$ and related to the doping
$x=1-n$. In the inset we show the square of the normalized plasma frequency 
as a function of temperature $T$ obtained in Ref.\cite{ioffe}
for YBa$_2$Cu$_3$O$_{6.6}$
}
\vskip 0.5cm

\caption{The density of states (DOS) $N(\omega)$ as a function of 
energy $\omega$ including both DDW order and superconductivity.
For the solid curve in each frame, the DDW gap $W_{0}=0.04$ and
the superconducting gap $\Delta_{0}=0.02$ while the dot-dashed
curve in the top frame has $W_{0}=0.03$ instead. The frames
differ in choice of chemical potential $\mu$.
In the top frame $\mu=-0.25$, middle $\mu=-0.08$, and in the bottom
$\mu=0$. The dahed line indicates $\omega=|\mu|$ and FS stands for
the Fermi surface.
}

\end{figure}


\begin{references}

\bibitem{timusk} T. Timusk and B. Statt, Rep. Prog. Phys. {\bf 62},
61 (1999) and references therein.

\bibitem{basov} D. N. Basov, S. I. Woods, A. S. Katz, E. J. Singley,
R. C. Dynes, M. Xu, D. G. Hinks, C. C. Homes, and M. Strongin,
Science {\bf 283}, 49 (1999); 

\bibitem{katz} A. S. Katz, S. I. Woods, E. J. Singley,
T. W. Li, M. Xu, D. G. Hinks, R. C. Dynes, and D. N. Basov,
Phys. Rev. B {\bf 61}, 5930 (2000).

\bibitem{emery} V. J. Emery and S. A. Kivelson, Nature (London) {\bf 374},
434 (1995).

\bibitem{chen} Q. Chen, I. Kosztin, B. Janko, and K. Levin,
Phys. Rev. Lett. {\bf 81}, 4708 (1998).

\bibitem{chakravarty1} S. Chakravarty, R. B. Laughlin, D. Morr, and
C. Nayak, Phys. Rev. B {\bf 63}, 094503 (2001). 

\bibitem{han} J. H. Han, Q. -H. Wang, and D. -H. Lee,
cond-mat/0012450; Q. -H. Wang, J. H. Han, and D. -H. Lee,
cond-mat/0102048.

\bibitem{tewari} S. Tewari, H. -Y. Kee, C. Nayak, and
S. Chakravarty, cond-mat/0101027.

\bibitem{chakravarty2} S. Chakravarty, H. -Y. Kee, and C. Nayak,
cond-mat/0101204.

\bibitem{cooper} S. L. Cooper and K. E. Gray in
{\it Physical properties of high temperature
superconductors V} edited by D. M. Ginsberg (World Scientific).

\bibitem{takenaka} K. T. Takenaka, K. Mizuhashu, H. Takagi, and
S. Uchida, Phys. Rev. B {\bf 50}, 6534 (1994).

\bibitem{forro} L. Forro, V. Ilakovac, J. R. Cooper, C. Agache,
and Y. Henny, Phys. Rev. B {\bf 46}, 6626 (1992).

\bibitem{fgt} R. A. Ferrell and R. E. Glover, Phys. Rev. {\bf 109},
1398 (1958); M. Tinkham and R. A. Ferrell, Phys. Rev. Lett. {\bf 2},
331 (1959).

\bibitem{ioffe} L. B. Ioffe and A. J. Millis, Science {\bf 285}, 1241 (1999).

\bibitem{zhu} J. -X. Zhu, W. Kim, S. C. Ting, and J. P. Carbotte,
cond-mat/0105580.

\bibitem{tallon} J. L. Tallon and J. W. Loram, cond-mat/0005063.

\bibitem{chakravarty} S. Chakravarty, H. -Y. Kee, and E. Abrahams,
Phys. Rev. Lett. {\bf 82}, 2366 (1999).

\bibitem{kim1} W. Kim and J. P. Carbotte, Phys. Rev. B {\bf 61},
R11886 (2000).


\bibitem{homes} C. C. Homes, T. Timusk, D. A. Bonn, R. Liang, and
W. N. Hardy, Physica C{\bf 254}, 265 (1995).

\bibitem{renner} Ch. Renner, B. Revaz, J. Genoud, K. Kadowaki,
and O. Fischer, 
Phys. Rev. Lett. {\bf 80}, 149 (1998).

\bibitem{dewilde} Y. DeWilde, 
N. Miyakawa, P. Guptasarma, M. Iavarone, L. Ozyuzer,
J. Zasadzinski, P. Romano, 
D. Hinks, C. Kendziora, G. Crabtree, and K. Gray, 
Phys. Rev. Lett. {\bf 80}, 153 (1998)

\bibitem{miyakawa} N. Miyakawa, P. Guptasarma, J. Zasadzinski,,
D. Hinks, and K. Gray, Phys. Rev. Lett. {\bf 157} (1998).

\bibitem{krasnov1}  V. M. Krasnov, A. Yurgens, D. Winkler,
P. Delsing, and T. Claeson,
Phys. Rev. Lett. {\bf 84},
5860 (2000).

\bibitem{krasnov2} V. M. Krasnov, A. E. Kovalev, A. Yurgens, and D. Winkler,
Phys. Rev. Lett. {\bf 86}, 2657 (2001).

\bibitem{shibauchi} See also new data of
T. Shibauchi, L. Krusin-Elbaum, M. Li, M. P. Maley, and
P. H. Kes, cond-mat/0104261.

\end{references}
\end{document}